# DVFS as a Security Failure of TrustZone-enabled Heterogeneous SoC

El Mehdi BENHANI, Lilian BOSSUET

Hubert Curien Laboratory
University of Lyon, Saint-Etienne, France
Emails: elmehdi.benhani@univ-st-etienne.fr,
lilian.bossuet@univ-st-etienne.fr

*Abstract*—Today, most embedded systems use Dynamic Voltage and Frequency Scaling (DVFS) to minimize energy consumption and maximize performance. The DVFS technique works by regulating the important parameters that govern the amount of energy consumed in a system, voltage and frequency. For the implementation of this technique, the operating system (OS) includes software applications that dynamically control a voltage regulator or a frequency regulator or both. In this paper, we demonstrate for the first time a malicious use of the frequency regulator against a TrustZone-enabled System-on-Chip (SoC). We use frequency scaling to create a covert channel in a TrustZone-enabled heterogeneous SoC. We present three different attacks, the first is discreet transmission of sensitive data from the SoC to outside, using electromagnetic emission. The second attack is the inside-SoC transfer of valuable data from a secure ARM core to a non-secure one. The last attack is the inside-SoC transfer of data between a non-trusted third party IP embedded in the programmable logic part of the SoC and a processor core.

*Index Terms*— ARM TrustZone, Embedded system security, AXI bus, Hardware Trojan, DVFS.

## I. INTRODUCTION

Systems-on-Chip (SoC) are becoming increasingly complex as they integrate many functionalities including third party IPs, which raises awareness of the need to protect the SoC from security failure. Today, one of the significant threats facing SoC is the covert channel transmission of valuable data. This security attack allows the attacker to transfer data between processes that are not authorized by the security policy to communicate. In general, a covert channel transmission uses an intruder process that transfers valuable information to a receiver process that decodes it, and uses it for malicious purposes. Many methods to create covert channels can be found in the literature, but most use shared resources such as cache memory [1].

Most modern SoC are equipped with Dynamic Voltage and Frequency Scaling (DVFS) capability to reduce power consumption and maximize performance. DVFS is a framework that makes it possible to change the frequency and/or operating voltage of a processor based on system performance requirements at a given point in time. The framework uses kernel drivers to control the hardware frequency and/or voltage regulator.

In this paper, we demonstrate for the first time a malicious use of the frequency regulator against TrustZone-enabled SoC.

We present three attacks implemented in the TrustZone-enabled SoC Xilinx Zynq-7010. All these attacks use frequency scaling to enable covert channel transmission.

We first present the related work in Section 2, followed by a description of the threat model in Section 3, the targeted platform, the prototype system, and the used protocol in Section 4, the three attacks in Section 5. Finally, we conclude the paper in Section 6.

## II. RELATED WORK

Covert channel attack is a well-known type of security attack in SoC. It is generally based on shared resources. Lipp et al. [1] presented a covert timing channel using a cache shared between two unprivileged processes. The covert communication is based on the famous cache attacks Evict+Reload, Flush+Reload, and Flush+Flush. The intruder process and receiver process use the time access to some addresses of a shared library to detect a cache hit or miss. This hit and miss is translated into a logical 0 and 1.

Masti et al. [2] demonstrated a thermal covert channel using the thermal sensor included in each processor in a multiprocessing platform. In their attack, the intruder process uses a core workload to heat the platform, thereby allowing a receiver process (core in the same platform) to decode the temperature variation as a logical 0 and 1.

Alagappan et al. [3] demonstrated a covert channel using frequency modulation. They used DVFS to transfer sensitive data between two cores that share the same clock. In their attack, the intruder process uses a core workload to affect the CPU frequency, which changes according to the CPU frequency governor mode used (performance, powersave, userspace, ondemand, conservative). The receiver process reads the frequency and translates it into a logical 0 or 1.

Like in [3], the attacks presented in this paper also use frequency modulation to send sensitive data between an intruder process and a receiver process. But unlike [3], the two processes have different security statuses in a TrustZone-enabled SoC. What is more, our attacks use direct modification of the register related to the frequency regulator. We also present for the first time a new covert communication in heterogeneous SoC, which is the communication between a hardware IP embedded in a *programmable logic* (FPGA fabric for example) and an ARM core.

This work was carried out in the framework of the FUIAAP20-Project TEEVA supported by Bpifrance.

## III. THREAT MODEL

In this paper, the general threat is that two processes prohibited from communicating with each other by the security policies, want to share information illegitimately. The two processes have a different security status in the TrustZone-enabled heterogeneous SoC. One secure process (intruder process) with access to some critical assets, and one non-secure process (receiver process) that is not allowed direct access to secure elements because of the memory management unit (MMU) rules and TrustZone protection [4]. We assume that the intruder process has write permission over the shared resource and that the receiver process has read permission.

## IV. PROTOTYPE SYSTEM

For our experiment, we used the Xilinx Zynq-7010 SoC, a TrustZone-enabled heterogeneous SoC. The Xilinx Zynq-7010 is compliant with TrustZone technology but the software and hardware implementation of the TrustZone security services involves a complex process. Interested readers can follow the cost free on-line tutorial [5] on designing a TrustZone-enable system with the Xilinx Vivado CAD tool.

Figure 1 shows the prototype system used for the experiment. In this prototype, SoC is partitioned into a *Processing System* (in blue in Figure 1), and a *Programmable Logic* (in yellow in Figure 1). The *Processing System* integrates a dual ARM core (Cortex-A9) that shares the same clock source (brown dashed line in Figure 1). The two cores also share an external memory (in white in Figure 1) with the *Programmable Logic*.

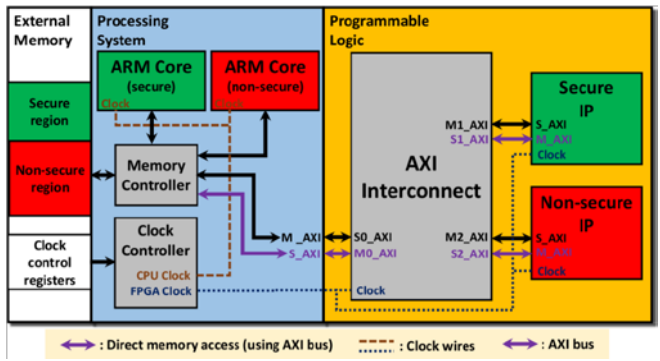

Fig. 1. Prototype system with Xilinx Zynq-7010 SoC

The TrustZone technology helps partition the external memory, the *Programmable Logic*, and the *Processing System* into secure (in green in Figure 1) and non-secure (in red in Figure 1) memory. The secure ARM core of the *Processing System* runs a custom trusted operating system that is stored in the secure region of the external memory. The second ARM core (non-secure) runs a general operating system that is stored in the non-secure region of the external memory. Both ARM cores share the same clock source from the SoC clock controller. The *Programmable Logic* includes a secure IP and non-secure IP. Both IPs have direct access to the entire external memory with no control by the ARM processor (Purple AXI line in Figure 1).

In the following Section, we describe the four attacks. To exchange data between the intruder process and the receiver, the attacks use a simple protocol that starts by sending a specific word like 0xAAAAAAAA, followed by the size of the data to be transferred, and the data. The transfer finishes by sending the same word as at the beginning. Like in the related work on covert channels, during the experiment, we did not focus on reaching the highest bandwidth.

## V. ATTACKS

This Section presents four attacks that make use of the DVFS covert channel. Figure 2 shows the attack paths of the four attacks.

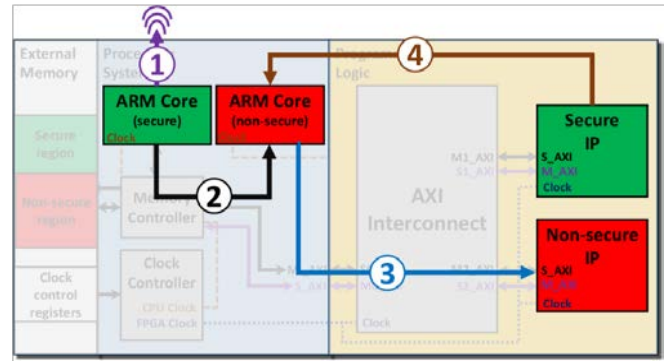

Fig. 2. Attack paths

### A. Attack #1

The first attack is the transfer of sensitive data from the secure ARM core to outside SoC using electromagnetic emission (purple attack path #1 in Figure 2). In 2015, Bossuet et al. [6] demonstrated that the electromagnetic channel is a powerful covert channel for discrete transmission from a SoC to outside. Nevertheless, in [6] an intruder circuit (spy circuit) was added to the design. Unlike [6], in the present paper, the attacker does not use an additional block for electromagnetic emission because the frequency modulation is done by the DVFS system directly. The attacker uses an electromagnetic probe and a real-time spectrum analyzer to decode the received data, as shown in Figure 3.

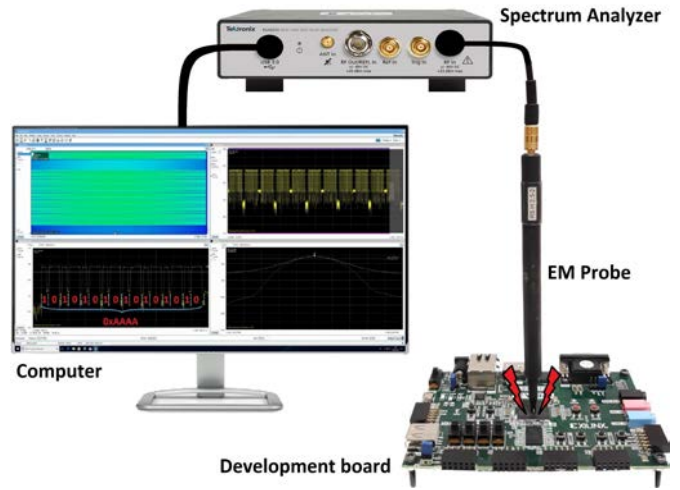

Fig. 3 Real-time spectral analysis of the electromagnetic leakage of secret (or sensitive) data

The attacker does not need to make an electromagnetic map of the studied SoC in order to detect the location of the targeted signal, a simple hand sweep is sufficient to reveal the position of a strong signal. This is due to the large number of wires connected to the manipulated clock source in this attack.

For this attack, the trusted operating system includes a malicious code in the driver that controls the frequency regulator. The malicious code uses frequency modulation to transfer the data, as presented in algorithm 1. This algorithm uses the same frequency *freq_1* to send a logical 1 and 0, and keeps this frequency for a long period of time to send a logical 1 *Tempo_1*, and for a short period to send a logical 0 *Tempo_2*. Between sending the two bits, the algorithm changes the CPU frequency *actuel_CPU_freq* to another frequency *freq_2*, and keeps it for a short period, *Tempo_3* in order to help the attacker distinguish between a logical 1 or 0.

*Algorithm* 1: *Frequency modulation*

**Input**: *data_to_transfer*
    **for** i = *data_to_transfer_size* to 0 **do**
        **if** (*data_to_transfer*[i] = 1) **then**
          *actuel_CPU_freq* = *freq_1*;
          loop for *Tempo_1*;
        **else**
          *actuel_CPU_freq* = *freq_1*;
          loop for *Tempo_2*;
        **end if**;
        *actuel_CPU_freq* = *freq_2*;
        loop for *Tempo_3*;
    **end for**;

This algorithm has many parameters that affect the size of the bandwidth: *freq_1*, *freq_2*, *Tempo_1*, *Tempo_2*, *Tempo_3*. Figure 4 presents two received data decoding for two different set of parameters.

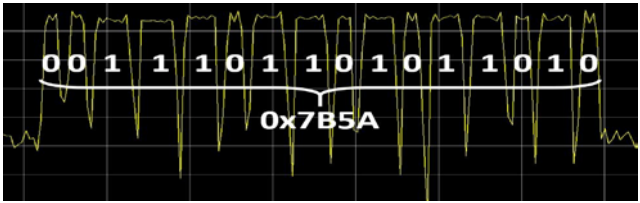

**a:** freq 1 = 325MHz, freq 2 = 433MHz,
Tempo 1 = 400, Tempo 2 = 200, Tempo 3 = 200

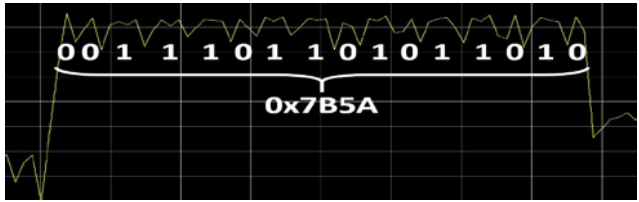

**b:** freq 1 = 325 MHz, freq 2 = 433 MHz,
Tempo 1 = 200, Tempo 2 = 100, Tempo 3 = 25

Fig. 4. Decoding received data, a - bandwidth = $1,42.10^5$ bps, b - bandwidth = $3,33.10^5$ bps

Figure 4 shows the relation between the temporal parameters (*Tempo_1*, *Tempo_2*, *Tempo_3*) and the bandwidth. Indeed, for high tempo values (Figure 4a), it is simple to decode the received data directly on the screen, but the bandwidth is smaller.

*B. Attack #2*

The second attack is a transfer of sensitive data from the secure ARM core to the non-secure one (black attack path #2 in Figure 2). In the Xilinx Zynq-7010 SoC, the two ARM cores are not well isolated because they are connected to the same clock. This attack uses this isolation issue to create a covert channel. It uses an intruder process included in the frequency regulator driver of the trusted operating system, and a receiver process included in the general operating system to decode the stolen data. The two processes use a direct read and/or write to the register related to the manipulated clock.

The intruder process uses algorithm 2 to transfer data. To send a logical 1, the algorithm uses the switch from *freq_1* to *freq_2*. To send a logical 0, the algorithm uses the switch from *freq_2* to *freq_1*. The algorithm holds the two frequencies for a short period of time. The method presented in the previous section also works, but if the transferred data are too long, th method has a high error ratio, and it is hard to synchronize the two cores using it. The method using algorithm 2 (rising and falling edge method, if we interpret the *freq_1* as low level and *freq_2* as high level) make it possible to reach $6.10^4$ bps in bandwidth and 0% in error ratio.

*Algorithm* 2: *Frequency modulation*

**Input**: *data_to_transfer*
    **for** i = *data_to_transfer_size* to 0 **do**
        **if** (*data_to_transfer*[i] = 1) **then**
          *actuel_CPU_freq* = *freq_1*;
          loop for *Tempo_1*;
          *actuel_CPU_freq* = *freq_2*;
          loop for *Tempo_1*;
        **else**
          *actuel_CPU_freq* = *freq_2*;
          loop for *Tempo_1*;
          *actuel_CPU_freq* = *freq_1*;
          loop for *Tempo_1*;
        **end if**;
    **end for**;

The receiver process uses algorithm 3 to decode the received data. At each *Tempo_sampling*, the code takes a sample by directly reading the clock register. If the algorithm detects a rising edge, it stores a logical 1 in the stolen data array, and if it detects a falling edge, it stores a logical 0 in the array. The choice of the sampling time *Tempo_sampling* is crucial to not miss any information. It should be smaller than the tempo *Tempo_1* used in the intruder process.

*C. Attack #3 and #4*

This section presents two attacks, one is a covert communication from the secure ARM core to the non-secure block IP (blue attack path #3 Figure 2), and one is a covert communication from the secure IP to the non-secure ARM core (brown attack path #4 Figure 2).

*a) From the secure ARM core to the non-secure IP*

This attack is a transfer of valuable data from the secure ARM core to the non-secure block IP. It uses a malicious code inserted in the driver of the DVFS frequency regulator as the intruder process, and the non-secure IP block as the receiver process.

The intruder process controls two of the four clocks that feed the *Programmable Logic* logical gates. All four clocks are limited to 250 MHz, and, to save energy, can only be activated for some clock cycles. The malicious code uses this activation characteristic to transfer data. To send a logical 1, it activates the first controlled clock for 10 cycles, and for a logical 0, it activates the clock for 5 cycles. Between two successive bits, the clock is off for a period that is proportional to the size of the code between the activation of two clock cycles. There is no constraint on choosing the number of clock cycles to activate. For example, the attacker can choose 3 clock cycles to send a logical 1, and 2 clock cycles to send a logical 0. The malicious code controls the second clock by changing it to the same frequency as the first source in order to help the block IP decode the received data.

The malicious non-secure block IP is connected to the two clock sources, and uses them to decode the received data. It uses a counter that starts once the first clock is active, and resets once the clock stops. The counter is incremented with each rising edge of the first clock. Figure 5 shows an example of a modulated signal at the top and decoded data at the bottom.

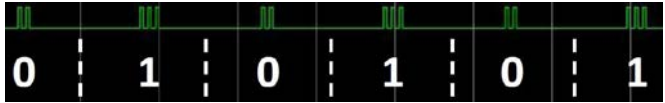

Fig. 5. Decoding received data

Table I lists four configurations of the benchmark used and the related bandwidth. The table shows that the size of the bandwidth is linked to the frequency and the activation cycles used. For this attack, the highest bandwidth reached is $125.10^6$bps.

TABLE I. BANDWIDTH ACCORDING TO THE FREQUENCY AND ACTIVATION CYCLES USED

| Frequency (MHz) | N° of cycles for a logical 1 | N° of cycles for a logical 0 | Bandwidth ($10^6$bps) |
|---|---|---|---|
| **250** | 10 | 5 | 50 |
| **250** | 3 | 2 | 125 |
| **100** | 10 | 5 | 20 |
| **100** | 3 | 2 | 50 |

*a) From secure IP to non-secure ARM core*

The last attack uses a malicious modification of the secure IP as the intruder process, and a code inserted in the general operating system as the receiver process. In [7], Benhani et al. present an example of this type of malicious block IP.

In this attack, the receiver process uses algorithm 3. The intruder process uses the direct memory access capability to modify the register related to the ARM core clocks. The malicious block IP does not know how the general operating system is mapped but by manipulating the clock connected to the ARM core, it can nevertheless transfer sensitive information. The intruder process uses the same method as the intruder process described in the second attack to send the data.

*Algorithm 3*: *Decoding data*

**Input**: *received_data_size*
**Output**: *stolen_data*
    **for** i = *received_data_size* to 0 **do**
        loop for *Tempo_sampling*;
        *last_freq* = *new_freq*;
        *new_freq* = read(*actuel_CPU_freq*);
        **if** (*last_freq = freq_1* and *new_freq = freq_2*) **then**
            *stolen_data*[i]= '1';
        **end if;**
        **if** (*last_freq = freq_2* and *new_freq = freq_1*) **then**
            *stolen_data*[i]= '0';
        **end if;**
    **end for;**
    **return** *stolen_data;*

## VI. CONCLUSION

In this paper, despite the security isolation provided by the TrustZone technology, we demonstrate the feasibility of using the frequency scaling used in modern SoC to enable covert channel transmission. The four attacks presented here successfully transferred sensitive data in a TrustZone-enabled SoC between an intruder process (secure) and a receiver process (non-secure) through malicious control of the frequency regulator. The paper also highlights the importance of the clock isolation in a SoC.

REFERENCES

[1] M. Lipp, D. Gruss, R. Spreitzer, C. Maurice, and S. Mangard, "Armageddon: Cache attacks on mobile devices."
[2] R. J. Masti, D. Rai, A. Ranganathan, C. Muller,̈ L. Thiele, and S. Capkun, "Thermal covert channels on multi-core platforms." in USENIX Security Symposium, 2015, pp. 865–880.
[3] M. Alagappan, J. Rajendran, M. Doroslovacki,̌ and G. Venkataramani, "DFS covert channels on multi-core platforms," in Very Large Scale Integration (VLSI-SoC), 2017 IFIP/IEEE International Conference on. IEEE, 2017, pp. 1 – 6.
[4] T. Alves, D. Felton, "TrustZone: Integrated Hardware and Soft-ware Security – Enabling Trusted Computing in Embedded Sys-tems," ARM white paper, 2004.
[5] E. M. Benhani, L. Bossuet, "Design a TrustZone-Enalble SoC usign Xilinx VIVADO CAD Tool," Technical Report, University of Lyon, 2017.
https://perso.univ-st-etienne.fr/bl16388h/VIVADO_TrustZone_tutorial.pdf
[6] L. Bossuet, P. Bayon, and V. Fischer, "An ultra-lightweight transmitter for contactless rapid identification of embedded ip in fpga," IEEE Embedded Systems Letters, vol. 7, no. 4, pp. 97–100, 2015.
[7] B. El Mehdi, C. Marchand, L. Bossuet, and A. Aubert, "On the security evaluation of the arm trustzone extension in a heterogeneous soc," in 30 th IEEE International System-on-Chip Conference, SOCC 2017. IEEE, 2017.